\documentclass[twocolumn,showpacs,preprintnumbers,amsmath,amssymb,prc]{revtex4}

\usepackage{graphicx}
\usepackage{dcolumn}
\usepackage{bm}
\usepackage{CJK}

\usepackage{color}

\begin{document}
\begin{CJK*} {GB} {gbsn}

\title{Electromagnetic field from asymmetric to symmetric heavy ion collision at 200 GeV/c}

\author {Yi-Lin Cheng}
\affiliation{ Shanghai Institute of Applied Physics, Chinese Academy of Sciences, Shanghai 201800, China}
\affiliation{ University of Chinese Academy of Sciences, Beijing 100049, China }

\author {Song Zhang} \thanks{Email: song\_zhang@fudan.edu.cn}
\affiliation{Key Laboratory of Nuclear Physics and Ion-beam Application (MOE), Institute of Modern Physics, Fudan University, Shanghai 200433, China}
\affiliation{ Shanghai Institute of Applied Physics, Chinese Academy of Sciences, Shanghai 201800, China}

\author {Yu-Gang Ma} \thanks{Email: mayugang@cashq.ac.cn}
\affiliation{Key Laboratory of Nuclear Physics and Ion-beam Application (MOE), Institute of Modern Physics, Fudan University, Shanghai 200433, China}\affiliation{ Shanghai Institute of Applied Physics, Chinese Academy of Sciences, Shanghai 201800, China}

\author {Jin-Hui Chen}
\affiliation{Key Laboratory of Nuclear Physics and Ion-beam Application (MOE), Institute of Modern Physics, Fudan University, Shanghai 200433, China}
\affiliation{ Shanghai Institute of Applied Physics, Chinese Academy of Sciences, Shanghai 201800, China}

\author {Chen Zhong}
\affiliation{Key Laboratory of Nuclear Physics and Ion-beam Application (MOE), Institute of Modern Physics, Fudan University, Shanghai 200433, China}
\affiliation{ Shanghai Institute of Applied Physics, Chinese Academy of Sciences, Shanghai 201800, China}

\date{\today}

\begin{abstract}
Electromagnetic fields produced in relativistic heavy-ion collisions are affected by the asymmetry of the projectile-target combination as well as the different initial configurations of the nucleus.
In this study, the results of the electric and magnetic fields produced for different combinations of ions, namely $^{12}$C + $^{197}$Au, $^{24}$Mg + $^{197}$Au, $^{64}$Cu + $^{197}$Au, and $^{197}$Au + $^{197}$Au at $\sqrt{s_{NN}} = 200$ GeV are demonstrated with a multi-phase transport model (AMPT). The configuration of the distribution of nucleons of $^{12}$C is initialized by a Woods-Saxon spherical structure, a three-$\alpha$-clustering triangular structure or a three-$\alpha$-clustering chain structure. It was observed that the electric and magnetic fields display different behavioral patterns for asymmetric combinations of the projectile and target nuclei as well as for different initial configurations of the carbon nucleus. The major features of the process are discussed.
\end{abstract}


\maketitle
\section{Introduction}

Extremely hot quark-gluon plasma (QGP) is produced in relativistic heavy-ion collisions, with large-scale collective motion at the partonic level for a short period of time and is found at the BNL-RHIC and CERN-LHC mega facilities~\cite{RHIC3}. The current focus in relativistic heavy-ion physics is on the determination of the QCD critical point from the regular hadronic matter to the QGP phase and the properties of the QGP~\cite {ref5, PBM, Luo, Song, Chen}. Previous research suggests that strong electromagnetic fields are produced in relativistic heavy-ion collisions~\cite {CM1} that may result in charge separation over the reaction plane similar to the chiral magnetic effect (CME)~\cite{CME1,Kharzeev}. 
A large number of theoretical research has been carried out to investigate the anomalous transport in heavy-ion collisions~\cite {C1, C2, Huang-Liao, Voloshin, PRC98-055201, PRC97-044906, PRC96-054909}. The CME is of interest as it may reflect the local parity and charge-parity violation in the case of strong interactions~\cite {CME2}. A few review articles on electromagnetic fields and anomalous transport in heavy-ion collisions are available in Refs.~\cite{RPP79-076302,AHEP2013} whereas the STAR~\cite{ref11,ref12}, PHENIX~\cite{ref13}, and ALICE~\cite{ref14} present the experimental aspects. Collaborative research on the charge-dependent two-particle correlation that corresponds qualitatively to the CME effect~\cite{ref15,ref16,ref17,ref18,ref19} was reported. In particular, the RHIC-STAR isobar runs in 2018 investigate the probability or percentage of the CME effect subtracted from the background by comparing the results of different isobar-colliding systems~\cite{CMEISO,CMEISO1,XuHJ,WangFQ,Zhao-new}.

In studies related to the CME, an estimation of the electromagnetic field strength is very important. Even though many simulations have been conducted for Au + Au and Pb + Pb systems, systematic calculations for different projectile-target combinations are not so abundant. This study presents the calculations of the electric and magnetic fields in  asymmetric to symmetric colliding systems, such as $^{12}$C + $^{197}$Au, $^{24}$Mg + $^{197}$Au, $^{64}$Cu + $^{197}$Au, and $^{197}$Au + $^{197}$Au. In particular, for the $^{12}$C + $^{197}$Au collision, the initial distribution of the nucleons in the carbon-12 nucleus is configured by three different geometrical distributions, namely,
 a three-$\alpha$-clustering chain structure, a three-$\alpha$-clustering triangular structure, or the Woods-Saxon nucleon distribution. A number of views on the $\alpha$-clustering structure of some specific nuclei were presented in theory and experiments~\cite{ref22,ref23,ref24,ref25,ref26,ref27,ref28,Ye,Aygun,XuZW,CaoXG}, however the effect on the calculation of the electromagnetic field in heavy-ion collisions was not included. It was observed in the present study that the electromagnetic field exhibits a dependence on the collision system, especially in the semi-central collisions, and has different values owing to the different nucleon configurations of the carbon nucleus for the $^{12}$C + $^{197}$Au system. These findings will contribute to some extent to an additional understanding of the CME phenomenon in different collision systems.

The paper is arranged as follows: In Section II, an introduction to the AMPT model and the algorithms for the $^{12}$C-clustering structure,  participant plane reconstruction in heavy-ion collisions, and calculation of the electromagnetic field are presented. The results and discussion of the effect of asymmetric nucleus-nucleus collision as well as the clustering configuration on the electromagnetic field are stated in Section III. Finally, a summary is presented in Section IV.

\section{Model and Algorithms}
\subsection{AMPT model}
A multi-phase transport model (AMPT)~\cite{ref29} was employed in the calculation that is composed of multiple processes to describe relativistic heavy-ion collisions, namely, 
the initial conditions simulated by the HIJING model ~\cite{HIJING}, the partonic interactions described by the ZPC model~\cite{ZPC}, the hadronization process by the Lund string fragmentation or coalescence model, and the hadronic rescattering process by the ART model~\cite{ART1,ART2}. 
In the HIJING model, the distribution of the nucleons of the two nuclei in a head-to-head collision is expressed by the Woods-Saxon distribution with momentum in the ($z$) direction, i.e., the direction of the beam.
In the overlapping region of the two colliding nuclei, minijet partons and soft string excitations are produced and
the initial coordinates and momentum distribution of these were obtained from the HIJING model and applied to calculate the electromagnetic field.

As mentioned above, the initialization of the nucleon distribution for the projectile and target is simulated by the Woods-Saxon distribution~\cite{HIJING}  that describes the distribution of the Mg, Cu, and Au nuclei in this study. However, several theoretical predictions were made for $^{12}$C on its possible $\alpha$-clustering configuration.
For example, a triangle-like configuration in the ground state was predicted by the fermionic molecular dynamics~\cite{chernykh2007structure} and the antisymmetrized molecular dynamics~\cite{Kanada-Enyo:2012yif} which was supported by experiment ~\cite{Marin-Lambarri:2014zxa}; a
three-$\alpha$ linear-chain configuration was also predicted as an excited state in the time-dependent Hartree-Fock theory~\cite{Umar:2010ck} and other approaches~\cite{Morinaga1966On}.
In concurrence with the above predictions coupled with the traditional Woods-Saxon structure, the nucleon configuration of $^{12}$C was initialized by three cases: (a) the three-$\alpha$ clusters in a chain structure, (b) the three-$\alpha$ clusters in a triangular structure, and (c)  the Woods-Saxon distribution of the nucleons from the HIJING model~\cite{HIJING}.

\subsection{$^{12}$C-clustering structure}

The radial center ($r_{\alpha}$) of the $\alpha$ clusters in $^{12}\mathrm{C}$ has a Gaussian distribution, $e^{-0.5\left(\frac{r_{\alpha}-r_c}{\sigma_{r_c}}\right)^{2}}$, where $r_c$ is the distribution center, $\sigma_{r_c}$ is the width of the distribution,
and the nucleons inside each $\alpha$ cluster are given by the Woods-Saxon distribution.
The parameters of $r_c$ and $\sigma_{r_c}$ can be obtained from the EQMD calculation~\cite{ref27,ref28}.
For the triangular structure, $r_c = $ 1.8 fm and $\sigma_{r_c}  = $ 0.1 fm;
For the chain structure, $r_c = $ 2.5 fm, $\sigma_{r_c}  = $ 0.1 fm for two $\alpha$ clusters, whereas the other cluster will be at the center in $^{12}\mathrm{C}$.
After the determination of the radial center of the $\alpha$ cluster, the centers of the three clusters are placed in an equilateral triangle for the triangular structure or along a line for the chain structure.

\subsection{Participant plane}
As is well known, the impact parameter $\vec{b}$ is defined along the direction of the center of the projectile and target nuclei, which is perpendicular to the beam direction $z$ and an event plane can be constructed by the beam direction $z$ and impact parameter.
In the AMPT model, the event plane angle $\Psi_{EP}$ is random, therefore the coordinate plane of every event was rotated to the same event plane, similar to the experimental method.
In the calculation, the participant plane angle $\Psi_{PP}$ is used to describe the event plane angle approximately as done in the references~\cite{ref24, C2, PartPlane-1,PartPlane-2,PartPlane-3}, and is defined by,
\begin{eqnarray}
\Psi_n\{PP\} = \frac{\tan^{-1}\left(\frac{\left<r_{part}^2\sin\left(n\phi_{part}\right)\right>}{\left<r_{part}^2\cos\left(n\phi_{part}\right)\right>}\right)+\pi}{n},
\label{PartPlaneDef}
\end{eqnarray}
where, $\Psi_n\{PP\}$ is the $n$-th order participant plane angle ( $n$ = 2 in this case), $r_{part}$ and $\phi_{part}$ are the coordinate position and azimuthal angle of the participants in the collision zone in the initial state, respectively, and the average $\left<\cdots\right>$ denotes the density weighting.

\subsection{Electromagnetic field algorithm}
\begin{figure*}[htbp]
	\includegraphics[width=18cm]{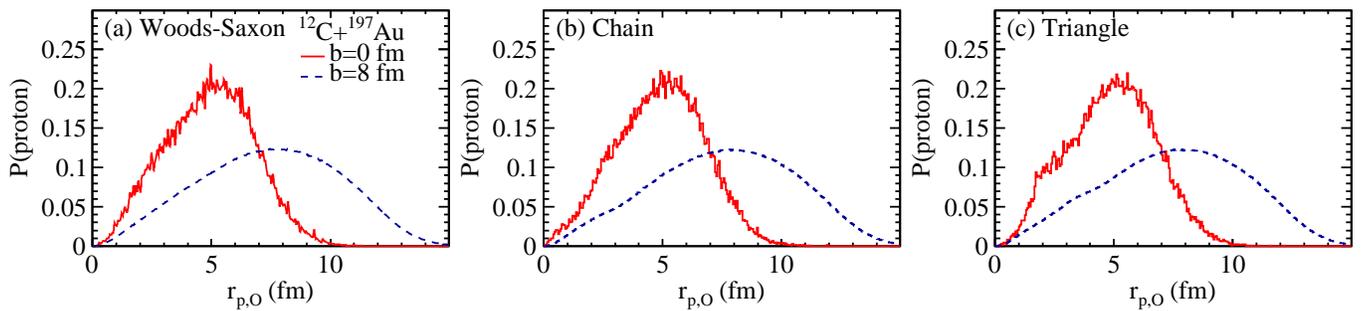}
	\caption {Proton distribution probability in central (at b = 0 fm) and peripheral (at b = 8 fm) $^{12}$C + $^{197}$Au collisions due to different configurations of $^{12}$C, (a) Woods-Saxon, (b) Chain, and (c) Triangular. }
	\label{figPProton}
\end{figure*}

In the calculation in this model, the origin of the coordinate system ($\vec{r}=0$) coincides with the centre of the collision zone. 
In this study, the electromagnetic field is calculated at the field point ($\vec{r}$=0, t=0) and the initial time t = 0 is defined as the moment when two colliding nuclei overlap completely. The proton distribution relative to the field point ($\vec{r}=0$) must be considered to avoid the 
divergence of the electromagnetic field (in the order of $e^2/r^2$) as $r$ tends to zero. Figure~\ref{figPProton} presents the probability of the proton distribution (P(proton)) as a function of $r_{p,O}$, which is the distance between the proton (of $^{12}$C or $^{197}$Au) and field point ($\vec{r}=0$, $t$=0). P(proton) increases and then decreases with the increase in $r_{p,O}$, the value of $r_{p,O}$ at the peak of P(proton) in the peripheral collisions (at b = 8 fm) being larger than in the central collisions (at b = 0 fm). It can be seen that the different configurations of $^{12}$C show a similar trend. Furthermore, it has also been verified that there is a similarity with other collision systems. In all the cases, it is observed that P(proton) is negligible near $r_{p,O}$=0 ($\vec{r}=0$). In the following calculation, the coordinate is set at a cut-off length $r>0.6$ fm as an approximation to avoid the divergence of the fields. 

In this study, the $Li\acute{e}nard-Wiechert$ potential was applied to calculate the electromagnetic fields~\cite{F1}.

\begin{eqnarray}
\begin{split}
e\vec{E}(t,\vec{r})=\frac{e^{2}}{4\pi}\sum_{n}Z_{n}\frac{\vec{R}_{n}-R_{n}\vec{v}_{n}}{({R_{n}-\vec{R}_{n}\cdot\vec{v}_{n}})^{3}}(1-v_{n}^{2}),\\
e\vec{B}(t,\vec{r})=\frac{e^{2}}{4\pi}\sum_{n}Z_{n}\frac{\vec{V}_{n}\times\vec{R}_{n}}{({R_{n}-\vec{R}_{n}\cdot\vec{v}_{n}})^{3}}(1-v_{n}^{2}),
\end{split}
\end{eqnarray}
where $Z_{n}$ is the charge number of the $n$-th particle, $\vec{R}_{n} = \vec{r} - \vec{r}_{n}$, where $\vec{r}$ is the position of the field point, and $\vec{r}_{n}$ is the position of the $n$-th particle at the retarded time $t_{n} = t - |\vec{r} - \vec{r}_{n}|$ and $t_{n}<t$. The objective is to calculate the electromagnetic fields at the position $\vec{r}$=0 and time $t = 0$. $\vec{E}(0,0)$ and $\vec{B}(0,0)$ are then marked as $\vec{E}$ and $\vec{B}$ for brevity. $\vec{v}_{n}$ is the velocity of each nucleon, where $v_{x} = v_{y} = 0$, $v_{z}^{2} = 1- (2m_{N}/\sqrt{s})^{2}$, $m_{N}$ is the mass of the nucleon. As $v_{z}$ is close to the velocity of light, in practice, the Lorentz contraction is taken into consideration. To investigate the calculation stability of electromagnetic fields, the cut-off length was tuned in a range from $r >$ 0.3 fm to $r >$ 0.9 fm and it was found there was no appreciable changing of the results.  And then a cut-off $r>0.6 fm$ was assumed in this calculation to avoid the divergence of the field. The possible correction of the classical Maxwell field equations by quantum electrodynamics (QED) was discussed in reference~\cite{RPP79-076302}, and it was found that the quantum correction can only amend the final results by a few percent.

\section{Results and Discussion}

\subsection{$\langle E_x\rangle/m^2_{\pi}$ and $\langle -B_y\rangle/m^2_{\pi}$ of systems}

The dependence on the impact parameter of the electric fields in the x-direction, $\langle E_x\rangle/m^2_{\pi}$, in Au + Au, Cu + Au, Mg + Au, and C + Au collisions are shown in Fig.~\ref{rot-Ex}. In each collision system,
$\langle E_x\rangle/m^2_{\pi}$ first shows an increasing trend and then decreases with a peak at a certain value of the impact parameter.

 Furthermore, it is seen that the values of $\langle E_x\rangle/m^2_{\pi}$ increase with increasing asymmetry between the projectile and target nuclei, i.e., from the most symmetric system of Au + Au, to  Cu + Au, to Mg + Au, and to the most asymmetric C + Au collision system. The inset of Fig.~\ref{rot-Ex} shows the $\langle E_x\rangle/m^2_{\pi}$ as a function of the impact parameter in the C + Au collisions where $^{12}$C is  configured by different initial geometries, namely, a three-$\alpha$ chain or triangular structure, or the Woods-Saxon nucleon distribution. The dependence on the impact parameter is similar for the three configurations. In the central collisions (small impact parameters) $\langle E_x\rangle/m^2_{\pi}$ presents similar values,  whereas in the peripheral collisions, $\langle E_x\rangle/m^2_{\pi}$ is larger for the Woods-Saxon distribution than for the other two cases, being the same for the chain and triangular structures. 
 It is interesting to note that in the semi-peripheral collisions $\langle E_x\rangle/m^2_{\pi}$ emerges as a peak and has the lowest value in the case of the chain structure. In this range of the impact parameter, the electromagnetic effect is always significant.

 The magnetic field $\langle -B_y\rangle/m^2_{\pi}$ has a similar dependence on the impact parameter but the value of $\langle -B_y\rangle/m^2_{\pi}$ decreases with increasing asymmetry between the projectile and target nuclei as shown in Fig.~\ref{rot-By}. The dependence of $\langle -B_y\rangle/m^2_{\pi}$ on the collision system is consistent with the results of reference~\cite{F1,ref15}. From figure~\ref{rot-Ex} and figure~\ref{rot-By}, it is implied that the asymmetric projectile and target nuclear collisions produce a stronger electric field than the symmetrical collision system, but the magnetic field exhibits a reverse trend. In other words, a dominant effect of the electric and magnetic field is evident in asymmetrical and symmetrical collision systems, respectively. In addition, the value of $\langle -B_y\rangle/m^2_{\pi}$ for C + Au with the three-$\alpha$ $^{12}$C chain configuration is slightly larger than that with either the Woods-Saxon nucleon distribution or the triangular structure, as shown in the inset of figure~\ref{rot-By}.

\subsection{Decomposition of $\langle E_x\rangle/m^2_{\pi}$ and $\langle -B_y\rangle/m^2_{\pi}$ into projectile and target sides}

\begin{figure}[htbp]
	\includegraphics[width=9cm]{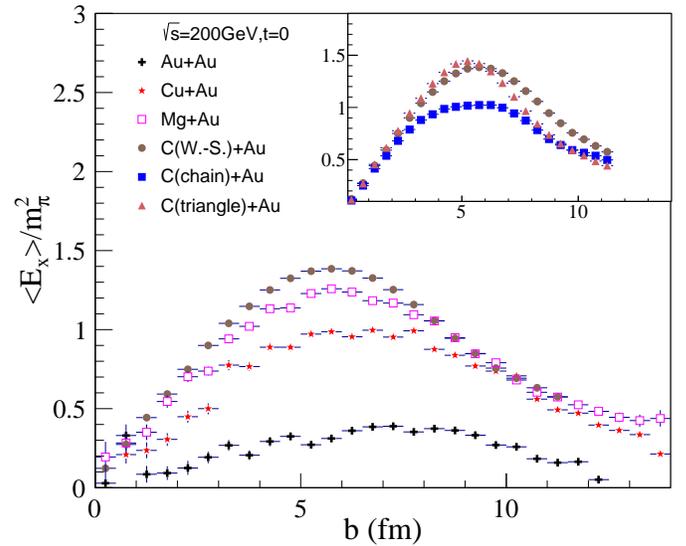}
	\caption {Impact parameter dependence of the $x$-component of electric field ($\langle E_x\rangle/m_\pi^2$) in the collision systems Au + Au, Cu + Au, Mg + Au, and C + Au. Inset displays $\langle E_x\rangle/m_\pi^2$ for different initial configurations of $^{12}$C.}
	\label{rot-Ex}
\end{figure}

\begin{figure}[htbp]
	\centering
	\includegraphics[width=9cm]{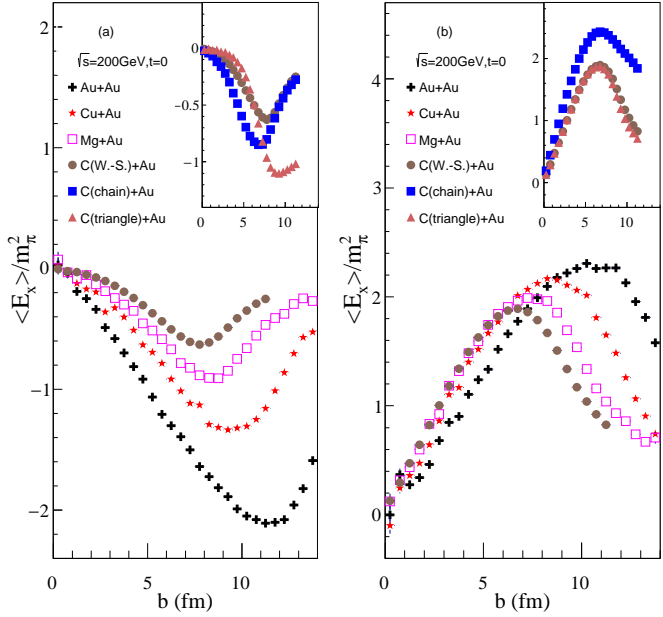}
	\caption {Impact parameter dependence of the $x$-component of electric field ($\langle E_x\rangle/m_\pi^2$) of projectile and target nuclei of Au + Au, Cu + Au, Mg + Au, and C + Au. The panel (a) is for the projectile and (b) is for the target. The insets are for $^{12}$C + Au for different initial configurations.}
	\label{ptEX}
\end{figure}

\begin{figure}[htbp]
	\centering
	\includegraphics[width=9cm]{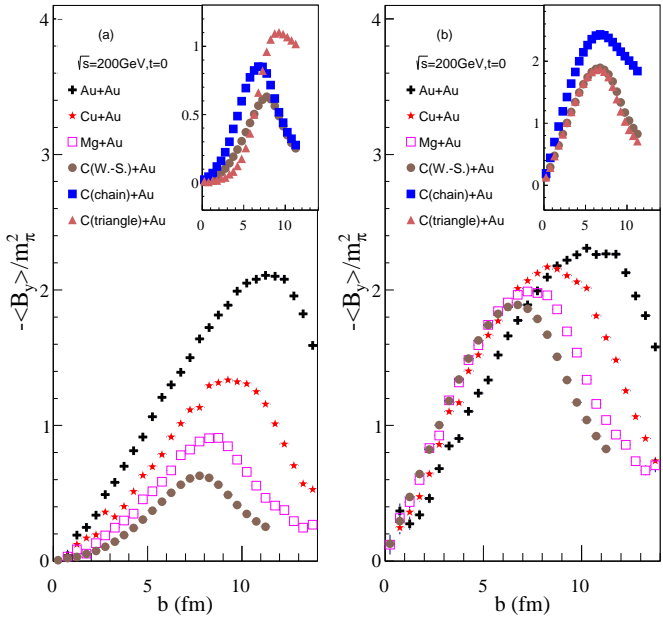}
	\caption {Same as Fig. 3 but for the opposite $y$-component of magnetic field (-$\langle B_y\rangle/m_\pi^2$) of projectile and target nuclei as shown in panel (b) and (b) respectively.
	}
	\label{pt-By}
\end{figure}

The dependence of the electromagnetic fields on the asymmetric collision system is further investigated by the fields generated by the projectile and target nucleons (protons). Fig.~\ref{ptEX} and Fig.~\ref{pt-By} present the electromagnetic fields $\langle E_x\rangle/m^2_{\pi}$ and $\langle -B_y\rangle/m^2_{\pi}$ produced on the projectile and target sides, respectively. The direction of movement of the projectile nucleus is parallel to the z-axis while that of the target nucleus is in the opposite direction. The panel (a) of Fig.~\ref{ptEX} shows the $\langle E_x\rangle/m^2_{\pi}$ produced by the projectile nucleus in different collision systems, namely the Au, Cu, Mg, and C nuclei, and the panel (b) shows those by the target nucleus (only Au nucleus). The electric field strength on the projectile side is negative and has a monotonic charge number dependence, i.e., the larger the proton number, the stronger the electric field. However, the electric field strength on the target side is positive and has a weak dependence on the proton number of the projectile, except for the peak position at a certain impact parameter.
Therefore, the total electric field contributed by the projectile and target nuclei depends on the collision system as shown in Fig.~\ref{rot-Ex}. 

\begin{figure}[htbp]
	\centering
	\includegraphics[width=9cm]{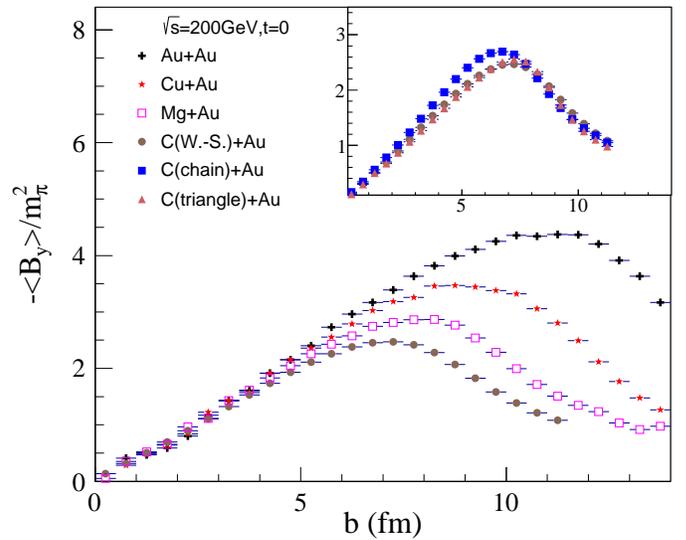}
	\caption {Same as Fig. 1 but for the opposite $y$-component of magnetic field (-$\langle B_y\rangle/m_\pi^2$).
	}
	\label{rot-By}
\end{figure}

The magnetic field generated by the projectile and target nuclei is shown in panel (a) and (b) of Fig.~\ref{pt-By} respectively. 
Unlike the electric field, the $\langle -B_y\rangle/m^2_{\pi}$ produced by the projectile and target nuclei has the same sign and the total contribution to the magnetic field is as shown in Fig.~\ref{rot-By}. However, with respect to the dependence on the charge number, the behavior on the projectile and target sides is the same as that of the corresponding electric field. 
Although the electric and magnetic fields have a similar dependence on the system on the projectile and target sides, the contribution from the overlapping region of the projectile and target nuclei leads to the results shown in Fig.~\ref{rot-Ex} and Fig.~\ref{rot-By}.  
It can be seen that the electric field is stronger in asymmetric collision systems, such as the C + Au collision, than in symmetric collision systems, which is unlike the trend shown by the magnetic field.  

The insets of figure~\ref{ptEX} and figure~\ref{pt-By} display the electric and magnetic fields produced by the projectile and target nuclei with different carbon configurations for the $^{12}$C + Au collision. On the target side (insets in panel (b)), it is seen that the three-$\alpha$ chain structure has a stronger EM field contribution, whereas for the three-$\alpha$ triangular structure and Wood-Saxon distribution it is the same.  

All the above results demonstrate that the dependence of the electromagnetic fields on the system originates from the competition of the fields between the projectile and target, and suggests that one can choose different collision systems to optimize the effect of either of the fields in experiments.

\subsection{$\langle E_y\rangle/m^2_{\pi}$ and $\langle B_x\rangle/m^2_{\pi}$ of  systems}

\begin{figure}[htbp]
	\centering
	\includegraphics[width=9cm]{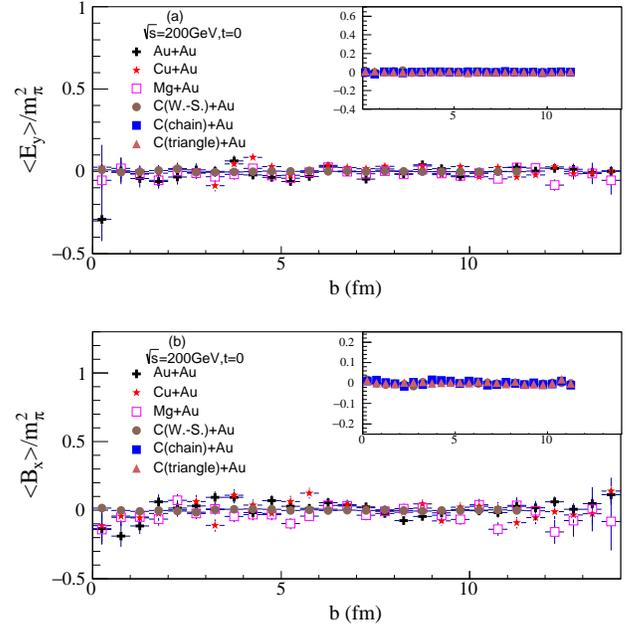}
	\caption {Impact parameter dependence of the $y$-component of electric field ($\langle E_y\rangle/m_\pi^2$) in panel (a) and the $x$-component of magnetic field ($\langle B_x\rangle/m_\pi^2$) in panel (b) of Au + Au, Cu + Au, Mg + Au, and C + Au. The insets are for $^{12}$C + Au for different initial configurations.}
	\label{rot-Ey}
\end{figure}

 Furthermore, the $y$ component of the electric field and $x$ component of the magnetic field, namely $\langle E_y\rangle$ and $\langle B_x\rangle$, are also calculated in this model.
Figure~\ref{rot-Ey} shows a zero value for $\langle E_y\rangle/m^2_{\pi}$ and $\langle B_x\rangle/m^2_{\pi}$ for all the impact parameters. Even for different configurations of carbon, as shown in the inset of figure~\ref{rot-Ey}, both $\langle E_y\rangle/m^2_{\pi}$ and $\langle B_x\rangle/m^2_{\pi}$ are zero. As the collision systems have been rotated event by event in this calculation, this action will result in a mirror symmetry of the collision geometry. Therefore, it is obvious to get a zero value of the $y$ component of the electric field and $x$ component of the magnetic field.

\subsection{$\langle E^2\rangle/m_\pi^4$ and $\langle B^2\rangle/m_\pi^4$}
\begin{figure}[htbp]
	\centering
	\includegraphics[width=9cm]{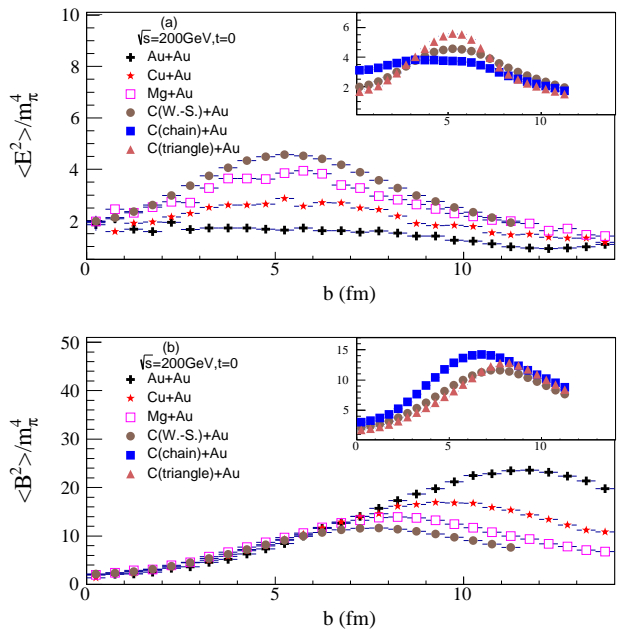}
	\caption {Impact parameter dependence of the square of the electric field ($\langle E^2\rangle/m_\pi^4$) in panel (a) and the square of the magnetic field ($\langle B^2\rangle/m_\pi^4$) panel (b) of Au + Au, Cu + Au, Mg + Au, and C + Au. The insets are for $^{12}$C + Au for different initial configurations.}
	\label{rot-Esq2}
\end{figure}

 The above results present the event averaged electromagnetic field excluding the fluctuations. Taking the fluctuation effect into consideration, $\langle E^2\rangle/m^4_{\pi}$ and $\langle B^2\rangle/m^4_{\pi}$ are calculated and shown in panel (a) and (b) of Fig.~\ref{rot-Esq2}, respectively. It is observed that only the x-component of the electric field and y-component of the magnetic field present non-zero values. The squared electromagnetic field strength is larger than $\langle E_x\rangle$ and $\langle -B_y\rangle$ as shown in figure~\ref{rot-Ex} and figure~\ref{rot-By}, even in the central collisions. This illustrates that the electromagnetic effect is more significant when the fluctuation of the fields is considered. The insets of figure~\ref{rot-Esq2} present the initial geometrical dependence of the electromagnetic fields. In the semi-peripheral collisions in particular,  the value of $\langle E^2\rangle$ in the triangular configuration of the carbon nuclei is the largest, followed by the Woods-Saxon configuration, whereas the chain configuration of the carbon nuclei has the smallest value. $\langle B^2\rangle$ in the chain structure is larger than in the other two configurations. This result suggests that the initial geometrical effect can be investigated by a system scan experiment of measurement of electromagnetic effects to understand the unusual nuclear structure besides the collective flow measurements proposed in references~\cite{ref22,ref24,ref26} in ultra-relativistic heavy-ion collisions.

\section{Summary}
\label{summary}
In summary, this study focused on the calculations of the electromagnetic fields for 
relativistic heavy-ion collision systems, from asymmetric to symmetric collisions, namely C + Au, Mg + Au, Cu + Au, and Au + Au at $\sqrt{s_{NN}}$ = 200 GeV, where $^{12}$C specifically had different initial configurations. The results demonstrated different behavioral patterns of the electromagnetic field for symmetric and asymmetric collision systems. The electric field was more significant in asymmetric than in symmetric collisions, whereas the magnetic field showed an opposite trend. This study elucidated different effects on the electric or magnetic fields produced in heavy-ion collisions. The initial geometrical effect of the unusual nuclear structure (i.e. $\alpha$-clusters in this study) was also studied and the electromagnetic field exhibited an initial geometrical dependence for different configurations of the carbon nucleus. Therefore, further research must be undertaken to understand the nuclear structure by investigating the electromagnetic effects through a system scan in relativistic heavy-ion collisions.

\begin{acknowledgments}
This work was supported in part by the National Natural Science Foundation of China under contract Nos. 11890710, 11890714, 11421505, 11875066, and 11775288, National Key R\&D Program of China under Grant No. 2016YFE0100900, the Key Research Program of Frontier Sciences of the CAS under Grant No. QYZDJ-SSW-SLH002, and the Key Research Program of the CAS under Grant NO. XDPB09.
\vspace{15cm}
\end{acknowledgments}

\end{CJK*}

\begin{thebibliography}{99}
	
\bibitem{RHIC3} J. Adams {\it et al.} (STAR Coll.), Nucl. Phys. A {\bf 757}, 102 (2005).
	
\bibitem{ref5} Cheuk-Yin Wong, Introduction to High-Energy Heavy-Ion Collisions, Singapore: World Scientific (1994)

\bibitem{Chen}J. Chen, D. Keane, Y. G. Ma, A. Tang, and Z. Xu, Phys. Rep. {\bf 760}, 1  (2018).

\bibitem{PBM}A. Andronic, P. Braun-Munzinger, K. Redlich, and J. Stachel,
Nature {\bf  561}, 321 (2018).

\bibitem{Luo}X.-F. Luo and N. Xu, Nuclear Science and Techniques  {\bf  28},  112 (2017).

\bibitem{Song}Huichao Song, You Zhou, K. Gajdosova,
Nuclear Science and Techniques {\bf 28}, 99 (2017).

	
\bibitem{CM1}  J. Rafelski  and B. M\"uller, Phys. Rev. Lett.  {\bf36}, 517 (1976).

\bibitem{CME1}  D. E. Kharzeev,  L. D. McLerran and H. J. Warringa, Nucl. Phys. A {\bf 227},   803 (2008).

\bibitem{Kharzeev}D.E. Kharzeev, J. Liao, S.A. Voloshin, G. Wang,
Prog.  Part. Nucl. Phys. {\bf 88}, 1 (2016).

\bibitem{C1}  Koichi Hattori, Xu-Guang Huang,  Nuclear Science and Techniques {\bf 28}, 26 (2017).


\bibitem{C2}  X. L. Zhao, Y. G. Ma, G. L. Ma,  Phys. Rev. C {\bf 97}, 024910 (2018)


\bibitem{Huang-Liao} Xu-Guang Huang and Jinfeng Liao, Phys. Rev. Lett. {\bf  110}, 232302 (2013).

\bibitem{Voloshin}Sergei A. Voloshin, Phys. Rev. Lett. {\bf  105}, 172301 (2010).

\bibitem{PRC98-055201} U. G\"ursoy, D. Kharzeev, E. Marcus, K. Rajagopal, and C. Shen, Phys. Rev. C {\bf 98} , 055201 (2018).
\bibitem{PRC97-044906} E. Stewart and K. Tuchin, Phys. Rev. C {\bf 97}, 044906 (2018).
\bibitem{PRC96-054909} V. Roy, S. Pu, L. Rezzolla, and D. H. Rischke, Phys. Rev. C {\bf 96}, 054909 (2017).

\bibitem{CME2} D. E. Kharzeev and Dam T. Son,  Phys. Rev. Lett. {\bf  106}, 062301 (2011).

\bibitem{RPP79-076302}X. G. Huang, Rep. Prog. Phys. {\bf 79}, 076302 (2016).
\bibitem{AHEP2013} K. Tuchin, Advances in High Energy Physics {\bf 2013}, 490495 (2013).


\bibitem{ref11} B. I. Abelev {\it et al.} (STAR Collaboration), Phys. Rev. C {\bf81}, 054908 (2010).

\bibitem{ref12} B. I. Abelev {\it et al.} (STAR Collaboration), Phys. Rev. Lett. {\bf103}, 251601 (2009).

\bibitem{ref13} N. N. Ajitanand, R. A. Lacey, A. Taranenko  {\it et al.}, Phys. Rev. C {\bf83}, 011901 (2011).

\bibitem{ref14} P.  Christakoglou, J. Phys. G {\bf38}, 124165 (2011).

\bibitem{ref15} W. T. Deng, X. G. Huang, Phys. Lett. B {\bf742}, 296 (2015).

\bibitem{ref16} G. L. Ma, X. G. Huang,  Phys. Rev. C {\bf91}, 054901 (2015).

\bibitem{ref17} A. Bzdak, V. Skokov,  Phys. Lett. B  {\bf710}, 171   (2012).

\bibitem{ref18} L. Ou, B. A. Li, Phys. Rev. C {\bf784}, 064605 (2011).

\bibitem{ref19} V. Voronyuk, V. D. Toneev, W. Cassing {\it et al.}, Phys. Rev. C {\bf83}, 054911 (2011).

\bibitem{CMEISO} L. Adamczyk {\it et al.} (STAR Collaboration), Phys. Rev. Lett. {\bf113}, 052302 (2014).

\bibitem{CMEISO1} B. Abelev {\it et al.}  (STAR Collaboration), Phys. Rev. Lett. {\bf110}, 012301 (2013).

\bibitem{XuHJ}Hao-jie Xu, Xiaobao Wang, Hanlin Li {\it et al.}, Phys. Rev. Lett. {\bf 121}, 022301 (2018).

\bibitem{WangFQ}F. Q. Wang, J. Zhao, Nuclear Science and Techniques  {\bf  29},  179 (2018).

\bibitem{Zhao-new}Xin-Li Zhao, Guo-Liang Ma, and Yu-Gang Ma, Phys. Rev. C {\bf 99}, 034903 (2019).

\bibitem{ref22}  W. Broniowski, E. R. Arriola, Phys. Rev. Lett. {\bf112}, 112501 (2014).

\bibitem{ref23}  M. Rybczynski, M. Piotrowska, W. Broniowski, Phys. Rev. C {\bf 97}, 034912 (2018).

\bibitem{ref24}  S. Zhang, Y. G. Ma, J. H. Chen, W. B. He, and C. Zhong, Phys. Rev. C  {\bf95}, 064904 (2017).

\bibitem{ref25}  A. Tohsaki, H. Horiuchi, P. Schuck, G. Ropke,  Phys. Rev. Lett. {\bf87}, 192501 (2001).

\bibitem{ref26} P. Bozek, W. Broniowski, E. R. Arriola {\it et al.}, Phys. Rev. C {\bf90}, 064902  (2014).

\bibitem{ref27} W. B. He, Y. G. Ma, X. G. Cao {\it et al.},  Phys. Rev. Lett. {\bf113}, 032506 (2014).

\bibitem{ref28} W. B. He, Y. G. Ma, X. G. Cao {\it et al.},  Phys. Rev. C {\bf94}, 014301 (2016).

\bibitem{Ye}Y. Liu, Y. L. Ye, Nuclear Science and Techniques  {\bf  29},  184 (2018).

\bibitem{Aygun} M.  Aygun, Z. Aygun, Nuclear Science and Techniques  {\bf 28}, 86 (2017).


\bibitem{XuZW} 	
Zhi-Wan Xu, Song Zhang, Yu-Gang Ma, Jin-Hui Chen, Chen Zhong, 
Nuclear Science and Techniques {\bf 29}, 186  (2018).

\bibitem{CaoXG}
X. G. Cao, E. J. Kim, K. Schmidt {\it  et al.}, Phys. Rev.  C {\bf 99}, 014606 (2019).

\bibitem{ref29} Z. W. Lin, C. M. Ko, B. A. Li {\it  et al.}, Phys. Rev. C {\bf 72}, 064901 (2005).

\bibitem{HIJING}   X.-N. Wang, M. Gyulassy, Phys. Rev. D {\bf44}, 3501 (1991).

\bibitem{ZPC} M. Gyulassy, X. N. Wang, Comput. Phys. Commun {\bf83}, 307 (1994).


\bibitem{ART1} B. A. Li, C. M. Ko, Phys. Rev. C {\bf52}, 2037 (1995).

\bibitem{ART2}B. A. Li, A. T. Sustich, B. Zhang {\it et al.}, Int. J. Mod. Phys. E  {\bf10}, 267 (2001).


\bibitem{chernykh2007structure}M. Chernykh, H. Feldmeier, T. Neff,  and P. von Neumann-Cosel, A. Richter,
Phys. Rev. Lett. {\bf 98}, 032501 (2007).

  \bibitem{Kanada-Enyo:2012yif}Y. Kanada-En'yo, M. Kimura, and A. Ono,     PTEP 2012, 01A202 (2012).

\bibitem{Marin-Lambarri:2014zxa} D. Marin-Lambarri, R. Bijker, M. Freer {\it et al.,}  Phys. Rev. Lett. {\bf 113}, 012502 (2014).

\bibitem{Umar:2010ck} A. S. Umar,  J. A. Maruhn,   N. Itagaki,  and  V. E. Oberacker,  Phys. Rev. Lett. {\bf 104}, 212503 (2010).

\bibitem{Morinaga1966On}H. Morinaga, Phys. Lett. {\bf 21}, 78 (1966).


\bibitem{PartPlane-1} Sergei A. Voloshin, Arthur M. Poskanzer,  Aihong Tang {\it et al.}, Phys. Lett. B {\bf 659}, 537 (2008).

\bibitem{PartPlane-2} B. Alver and G. Roland, Phys. Rev. C {\bf 81}, 054905 (2010).

\bibitem{PartPlane-3} Roy A. Lacey, Rui Wei, J. Jia {\it et al.}, Phys. Rev. C {\bf 83}, 044902 (2011).

\bibitem{F1} W. T. Deng, X. G. Huang, Phys. Rev. C {\bf 85}, 044907 (2012).

\end{thebibliography}
\end{document}